\documentclass[twocolumn, pra, a4paper, 10pt, superscriptaddress, nopacs]{revtex4-1}
\usepackage{siunitx}
\usepackage{graphicx}
\usepackage{xcolor}
\usepackage{physics}
\usepackage{nicefrac}
\usepackage{ulem}
\usepackage{float}
\usepackage[utf8]{inputenc}
\usepackage[colorlinks]{hyperref}

\newcommand{\CT}{CrTe\textsubscript{2} }
\newcommand{\hl}[1]{\textcolor{red}{#1}} 
\begin{document}
	
	\title{Characterization of room-temperature in-plane magnetization in thin flakes of \CT with a single spin magnetometer}
	
	\author{F. Fabre}
	\author{A. Finco}
	\affiliation{Laboratoire Charles Coulomb, Universit\'e de Montpellier and CNRS, 34095 Montpellier, France}
	\author{A. Purbawati}
	\author{A. Hadj-Azzem}
	\author{N. Rougemaille}
	\author{J. Coraux}
	\affiliation{Univ. Grenoble Alpes, CNRS, Grenoble INP, Institut NEEL, 38000 Grenoble, France}
	\author{I. Philip}
	\author{V. Jacques}
	\affiliation{Laboratoire Charles Coulomb, Universit\'e de Montpellier and CNRS, 34095 Montpellier, France}
	
	\date{\today}
	
	\begin{abstract}
We demonstrate room-temperature ferromagnetism with in-plane magnetic anisotropy in thin flakes of the \CT van der Waals ferromagnet. Using quantitative magnetic imaging with a single spin magnetometer based on a nitrogen-vacancy defect in diamond, we infer a room-temperature in-plane magnetization in the range of $M\sim 27$~kA/m for flakes with thicknesses down to $20$~nm.  In addition, our measurements indicate that the orientation of the magnetization is not determined solely by shape anisotropy in micron-sized \CT flakes, which suggest the existence of a non-negligible magnetocrystalline anisotropy. These results make \CT a unique system in the growing family of van der Waals ferromagnets, as it is the only material platform known to date which offers an intrinsic in-plane magnetization and a Curie temperature above $300$~K in thin flakes. 
	\end{abstract}

	\maketitle
	
\section{Introduction}
Ferromagnetic van der Waals (vdW) crystals offer numerous opportunities both for the study of exotic magnetic phase transitions in low-dimensional systems~\cite{Kosterlitz_1973} and for the design of innovative, atomically-thin spintronic devices~\cite{doi:10.1002/adma.201900065,doi:10.1002/andp.201900452}. Since the discovery of a two-dimensional (2D) magnetic order in monolayers of CrI$_3$~\cite{huangcri32017} and Cr$_2$Ge$_2$Te$_6$~\cite{gongcgt2017} crystals, the family of vdW ferromagnets has expanded very rapidly~\cite{Burch2018,Gong2019,Gibertini2019}. However, most of these compounds have a Curie temperature ($T_{\rm c}$) well below 300 K, which appears as an important drawback for future technological applications. An intense research effort is therefore currently devoted to the identification of high-$T_{\rm c}$ 2D magnets~\cite{doi:10.1002/andp.201900452}. 

In this context, the vdW crystal Fe$_{3}$GeTe$_{2}$ appears as a serious candidate because it can be grown in wafer-scale through molecular beam epitaxy and it exhibits a strong perpendicular magnetic anisotropy~\cite{FGTMBE,doi:10.1063/1.4961592}. Although its intrinsic $T_{\rm c}$ drops to $130$~K in the monolayer limit~\cite{FGT2018}, it might be raised above room-temperature either by ionic gating~\cite{Deng2018,WeberNanolett2019}, interfacial engineering~\cite{WangACSNano2020}, or by micro-patterning, as demonstrated so far for rather thick films~\cite{LiNanoLett2018}. In addition, other FeGeTe alloys, such as Fe$_4$GeTe$_2$ and Fe$_5$GeTe$_2$, exhibit high $T_{\rm c}$, still lower than room temperature but close to it~\cite{ACSNano2019,Seoeaay8912}. Another promising strategy consists in incorporating magnetic dopants into 2D materials to form dilute magnetic semiconductors~\cite{PhysRevLett.125.036802}. This approach was recently employed to induce room-temperature ferromagnetism in WSe$_2$ monolayers doped with vanadium~\cite{doi:10.1002/advs.201903076,arXivWse2}. Finally, an intrinsic ferromagnetic order was reported in epitaxial layers of VSe$_2$~\cite{Bonilla2018vse2}, MnSe$_{2}$~\cite{OHaramnse2018} and VTe$_{2}$~\cite{doi:10.1002/adma.201801043} under ambient conditions, although the interpretation of these experiments still remains debated~\cite{VSe2debate1,doi:10.1002/adma.201901185}.
	
In this work, we follow an alternative research direction by studying the room-temperature magnetic properties of micron-sized flakes exfoliated from a \CT crystal with $1T$ structure. In its bulk form, this layered transition metal dichalcogenide is a ferromagnet with {\it in-plane} magnetization, {\it i.e.} pointing perpendicular to the $c$ axis, and a $T_{\rm c}$ around $320$~K~\cite{Freitascrte22015}. This combination of properties is unique in the growing family of vdW ferromagnets. Recent studies have reported that the magnetic order is preserved at room temperature in exfoliated \CT flakes with thicknesses in the range of a few tens of nanometers~\cite{suncrte22019,Coraux2020}. However, obtaining quantitative estimates of the magnetization in such micron-sized flakes remains a difficult task, which requires the use of non-invasive magnetic microscopy techniques combining high sensitivity with high spatial resolution. These performances are offered by magnetometers employing a single nitrogen-vacancy (NV) defect in diamond as an atomic-size quantum sensor~\cite{Gupi_Nature2008,Maze2008,rondinmagnetometry2014}. In recent years, this microscopy technique has found many applications in condensed matter physics~\cite{casola_probing_2018}, including the study of chiral spin textures in ultrathin magnetic materials~\cite{TetienneNatComm2015,SkYacoby2018,ChauleauNatMat}, current flow imaging in graphene~\cite{YacobyNature2020} and the analysis of the magnetic order in vdW magnets down to the monolayer limit~\cite{Thielcri32019,Tetienne2020,arXivWrachtrup}. Here we use scanning-NV magnetometry to infer quantitatively the in-plane magnetization in exfoliated \CT flakes under ambient conditions. Our measurements confirm that the ferromagnetic order is preserved in few tens of nanometers thick flakes, although with a low room-temperature magnetization $M\sim 27$~kA/m. This value is five times smaller that the one measured in a bulk \CT crystal. Such a reduction of the magnetization is attributed to a decreased Curie temperature in exfoliated flakes. Moreover, our results show that shape anisotropy alone does not fix the in-plane orientation of the magnetization in micron-sized \CT flakes, pointing out the existence of a substantial magnetocrystalline anisotropy.

\section{Materials and methods}

\begin{figure}[t]
		\centering
		\includegraphics[scale=1.06]{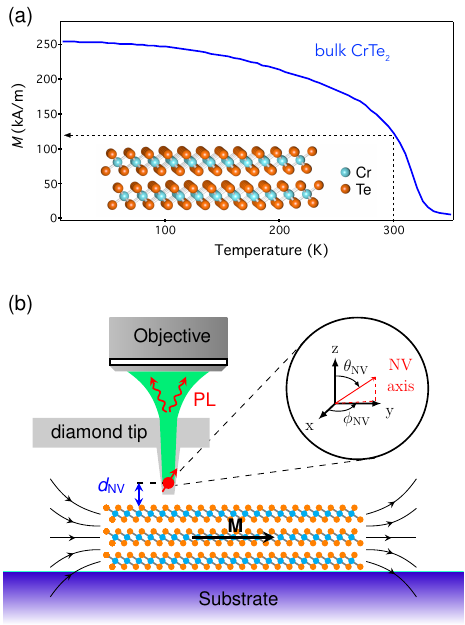}
		\caption{(a) Temperature dependence of the magnetization $M$ in a bulk crystal of \CT with $1T$ polytype. The measurement is performed by vibrating sample magnetometry under a magnetic field of $500$~mT. At room temperature, the magnetization is around $120$~kA/m (black dashed lines). The inset shows the layered crystal structure of $1T$-\CT. (b) A single NV defect (red arrow) localized at the apex of a diamond tip is scanned above exfoliated \CT flakes. A microscope objective placed above the tip is used to collect the photoluminescence (PL) of the NV defect under green laser excitation. In this work, the NV-to-sample distance is $d_{\rm NV}\sim 80$~nm and the quantization axis of the NV defect is characterized by the spherical angles ($\theta_{\rm NV}=58^{\circ},\phi_{\rm NV}=103^{\circ}$) in the laboratory frame of reference $(x,y,z)$. The black arrows indicate the magnetic field lines generated at the edges of a \CT flake with uniform in-plane magnetization $\mathbf{M}$.}
		\label{fig:intro_CrTe2}
	\end{figure}

A bulk $1T$-\CT crystal was synthesized following the procedure described in Ref.~\cite{Freitascrte22015}. The in-plane magnetization of this layered ferromagnet was first characterized as a function of temperature through vibrating sample magnetometry under a magnetic field of $500$~mT. The results shown in Fig.~\ref{fig:intro_CrTe2}(a) indicate a Curie temperature around $320$~K and a magnetization reaching $M\sim 120$~kA/m under ambient conditions. \CT flakes with thicknesses ranging from a few tens to a hundred of nanometers were then obtained by mechanical exfoliation and transferred on a SiO\textsubscript{2}/Si  substrate. We note that the probability to obtain thin \CT flakes through mechanical exfoliation is still very low compared to other layered transition metal dichalcogenides, such as MoS$_2$ or WSe$_2$~\cite{Coraux2020}. The thinnest flake studied in this work has a thickness of $\sim20$~nm. Like all van der Waals ferromagnets known to date, \CT flakes are unstable under oxygen atmosphere. However, a recent study combining X-ray and Raman spectroscopy has shown that oxidation of \CT flakes occurs typically within a day scale under ambient conditions and is limited to the very first outer layers~\cite{Coraux2020}. In this work, \CT flakes were not encapsulated and all the measurements were done within a day after exfoliation to mitigate oxidation. 

Magnetic imaging was performed with a scanning-NV magnetometer operating under ambient conditions~\cite{rondinmagnetometry2014}. As sketched in Fig.~\ref{fig:intro_CrTe2}(b), a single NV defect integrated into the tip of an atomic force microscope (AFM) was scanned above \CT flakes to probe their stray magnetic fields. At each point of the scan, a confocal optical microscope placed above the tip was used to monitor the magnetic-field-dependent photoluminescence (PL) properties of the NV defect under green laser illumination. In this work, we employed a commercial diamond tip (Qnami, Quantilever MX) with a characteristic NV-to-sample distance $d_{\rm NV}= 80\pm10$~nm, as measured through an independent calibration procedure~\cite{Hingant2015}. Two different magnetic imaging modes were used. In the limit of weak stray fields ($<5$~mT), quantitative magnetic field mapping was obtained by recording the Zeeman shift of the NV defect electron spin sublevels through optical detection of the electron spin resonance (ESR). This method relies on microwave driving of the NV spin transition combined with the detection of the spin-dependent PL intensity of the NV defect~\cite{Gruber1997}. For stronger magnetic fields ($>5$~mT), the scanning-NV magnetometer was rather used in all-optical, PL quenching mode~\cite{Gross_2018,Akhtar2019}. In this case, localized regions of the sample producing large stray fields are simply revealed by an overall reduction of the PL signal induced by a mixing of the NV defect spin sublevels~\cite{Tetienne_2012}. We note that the diameter of the scanning diamond tip is around $200$~nm in order to act as an efficient waveguide for the PL emission of the NV defect~\cite{Maletinsky2012,appeltips2016}. As a result, such tips cannot provide precise topographic information of the sample. For the thinnest flakes, the topography was thus imaged by using conventional, sharp AFM tips. 
	\begin{figure*}[t]
		\centering
		\includegraphics[scale=1.07]{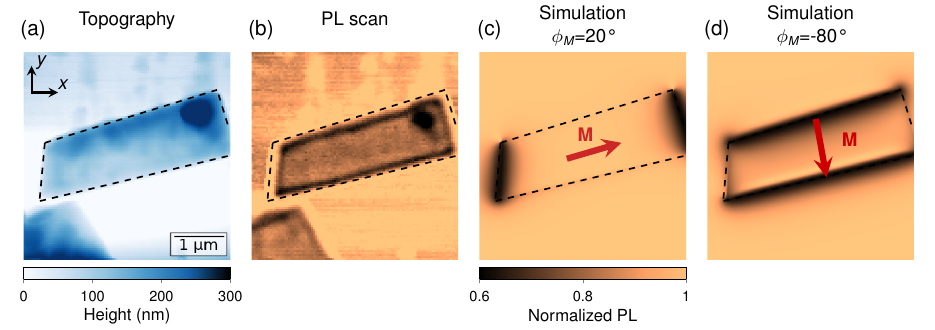}
		\caption{(a) AFM image of a \SI{150}{\nano\meter}-thick \CT flake recorded with the scanning-NV magnetometer. (b) Corresponding PL map showing a quenching contour at the edges of the flake. The shape of the flake is indicated by the black dashed lines. The recorded PL signal is normalized with the one measured far from the flake. The experiment is performed at zero external magnetic field. (c,d) Simulated PL maps for two different orientations $\phi_M$ of the in-plane magnetization: (c) $\phi_M=20^{\circ}$ and (d) $\phi_M=-80^{\circ}$. Here, the norm of the magnetization is fixed arbitrarily to its bulk value $M=120$~kA/m. The PL quenching induced by energy transfer between the NV sensor and the metallic sample is not included in the simulations.}
		\label{fig:quenching_CrTe2}
	\end{figure*}

For a ferromagnetic material, stray magnetic fields can be produced by magnetization patterns presenting a non-zero divergence ${\bf \nabla}\cdot{\bf M} \neq 0 $. Considering a \CT flake, this can occur (i) at the edges of the flake, (ii) at the location of non-collinear spin textures such as domain walls, and (iii) at the position of thickness steps on the flake.
In this work, the magnetization is inferred from the measurement of the stray field produced at the edges of the flake [Fig.~\ref{fig:intro_CrTe2}(b)]. Therefore, the analysis is made simpler for uniformly magnetized flakes with a homogeneous thickness. While uniform spin textures can be readily obtained by applying a bias magnetic field, obtaining thin \CT flakes with a uniform thickness through mechanical exfoliation is currently achieved with a low probability. Consequently, variations in thickness must be carefully taken into account when analyzing the experimental results.

\section{Results and discussion}

In a first experiment, we imaged a \CT flake with a large thickness $t \sim 150$~nm [Fig.~\ref{fig:quenching_CrTe2}(a)]. Considering the saturation magnetization of the bulk CrTe$_2$ crystal, magnetic simulations predict stray field amplitudes larger than $10$~mT at a distance $d_{\rm NV}\sim 80$~nm above the edges of the 150-nm-thick flake. The scanning-NV magnetometer was thus operated in the PL quenching mode for such a thick flake. Fig.~\ref{fig:quenching_CrTe2}(b) displays the resulting PL map. Several observations can be made. First, a quenching of the PL signal is observed when the NV defect is placed above the flake. This quenching effect, which is constant all over the flake, does not have a magnetic origin. It is linked to the metallic character of CrTe$_2$~\cite{PhysRevLett.95.063003,TislerNanolett2013}. Second, a stronger PL quenching is obtained at the flake edges, as expected for a single ferromagnetic domain. We note that the additional quenching spot observed at the top-right edge of the flake results from the stray field produced by a large variation of the thickness~[Fig.~\ref{fig:quenching_CrTe2}(a)]. Given the different sources of PL quenching involved in this experiment, quantitative estimates of the magnetization cannot be obtained. However, the analysis of the PL quenching distribution can be used to extract qualitative information about the orientation of the magnetization. To this end, simulations of the PL map were carried out considering only the effect of magnetic fields. The geometry of the flake used for the simulation was inferred from the AFM image simultaneously recorded with the NV microscope [Fig.~\ref{fig:quenching_CrTe2}(a)]. Assuming a uniform in-plane magnetization $\mathbf{M}$ with an azimuthal angle $\phi_{M}$ in the $(x,y)$ plane, the stray field was first calculated at a distance $d_{\rm NV}$ above the flake. The resulting PL quenching map was then simulated by using the model of magnetic-field-dependent photodynamics of the NV defect described in Ref.~\cite{Tetienne_2012}. Simulated PL maps obtained for two different orientations of the magnetization are shown in Figs.~\ref{fig:quenching_CrTe2}(c,d). In micron-sized flakes, shape anisotropy should favor a magnetization pointing along the long axis of the flake. Interestingly, the PL map simulated for such a magnetization direction disagrees with the experimental data [Fig.~\ref{fig:quenching_CrTe2}(c)], which instead suggest a magnetization pointing perpendicular to the long axis of the flake [Fig.~\ref{fig:quenching_CrTe2}(d)]. This result is a first indication of a non-negligible, in-plane magnetocrystalline anisotropy.

Thinner \CT flakes, {\it i.e.} producing less stray field, were then studied through quantitative magnetic field imaging. To this end, a microwave excitation was applied through a copper microwire directly spanned on the sample surface and the Zeeman shift of the NV defect's ESR frequency was recorded at each point of the scan. In the weak field regime, the ESR frequency is shifted linearly with the projection $B_{\rm NV}$ of the stray magnetic field along the NV defect quantization axis~\cite{rondinmagnetometry2014}. This axis was precisely measured by applying a calibrated magnetic field~\cite{Rondin2013}, leading to the spherical angles $\theta_{\rm NV}=58^{\circ}$ and $\phi_{\rm NV}=103^{\circ}$ in the laboratory frame $(x,y,z)$, as illustrated in Fig.~1(b). For these measurements, a bias field of \SI{3}{\milli\tesla} was applied along the NV defect axis in order to infer the sign of the stray magnetic field~\cite{rondinmagnetometry2014}. Furthermore, this bias field is strong enough to erase spin textures such as domain walls, given the weak coercitive field of \CT flakes~\cite{Coraux2020}.
\begin{figure*}[t]
		\centering
		\includegraphics[scale=1]{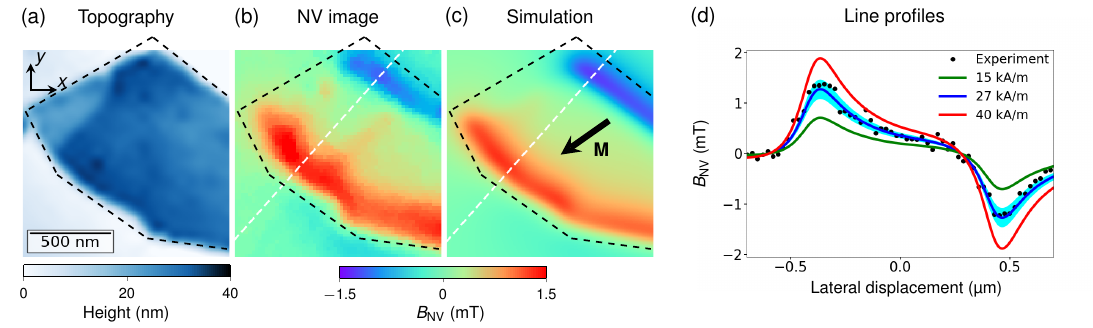}
		\caption{(a) AFM image of a \SI{35}{\nano\meter}-thick \CT flake. (b) Corresponding distribution of the magnetic field component $B_{\rm NV}$. (c) Simulated magnetic field distribution $B_{\rm NV}$ for a uniform in-plane magnetization $\mathbf{M}$ (black arrow) with an azimuthal angle $\phi_{M}=-145^{\circ}$ in the $(x,y)$ plane and a norm $M=27$~kA/m. The calculation is done at a distance $d_{\rm NV}=80$~nm above the flake, whose shape is extracted from the topography image [black dashed lined in (a)]. (d) Fitting of an experimental line profile (black markers) with the magnetic calculation (blue solid line). The position of the linecuts are indicated by the white dashed lines in (b) and (c). A magnetization $M=27 \pm 4$~kA/m is obtained, the uncertainty being illustrated by the blue shaded area. The red and green solid lines indicate the result of the calculation for $M=40$~kA/m and $M=15$~kA/m, respectively.}
		\label{fig:fullb_CrTe2}
	\end{figure*}
	
A map of $B_{\rm NV}$ recorded above a $35$-nm thick \CT flake is shown in Fig.~\ref{fig:fullb_CrTe2}(b). A stray magnetic field around $\pm 1.5$~mT is detected at two opposite edges of the flake. In principle, the underlying sample magnetization can be retrieved from such a quantitative magnetic field map by using reverse propagation methods with well-adjusted filters in Fourier-space~\cite{casola_probing_2018}. Under several assumptions, this method can be quite robust for magnetic materials with out-of-plane magnetization~\cite{Thielcri32019,Tetienne2020,arXivWrachtrup}. For in plane magnets, however, the reconstruction procedure amplifies noise and is thus much less efficient~\cite{Tetienne2020bis}. As a result, the recorded magnetic field distribution was rather directly compared to magnetic calculations in order to extract quantitative information on the sample magnetization. 
	
To obtain precise information on the geometry of the flake, the topography image was here measured with a sharp AFM tip [Fig.~\ref{fig:fullb_CrTe2}(a)]. The observed variations in thickness were included in the magnetic calculation (see Appendix A). Considering a uniformly magnetized flake with an azimuthal angle $\phi_{M}$, the stray field was calculated at a distance $d_{\rm NV}$ above the flake and then projected along the NV quantization axis in order to simulate a map of $B_{\rm NV}$. By comparing experimental data with magnetic maps simulated for different values of the angle $\phi_{M}$, the orientation of the in-plane magnetization was first identified, leading to $\phi_{M}=-145\pm5^{\circ}$, {\it i.e.} pointing along the short axis of the flake [Fig.~\ref{fig:fullb_CrTe2}(c)]. Once again, this result suggests the presence of a non-negligible magnetocrystalline anisotropy, in agreement with recent works~\cite{Coraux2020}. The norm $M$ of the magnetization vector was then estimated by fitting experimental line profiles with the result of the calculation, leading to $M=27 \pm 4$~kA/m~[Fig.~\ref{fig:fullb_CrTe2}(d)]. We note that the stray field amplitude depends on several parameters including $d_{\rm NV}$, $\phi_{M}$, the flake thickness $t$ and the NV defect quantization axis $\left(\theta_{\rm NV},\phi_{\rm NV}\right)$. Any imprecisions on these parameters directly translate into uncertainties on the evaluation of the magnetization $M$. A detailed analysis of uncertainties is provided in Appendix~B.

	\begin{figure}[t]
		\centering
		\includegraphics[width = 8cm]{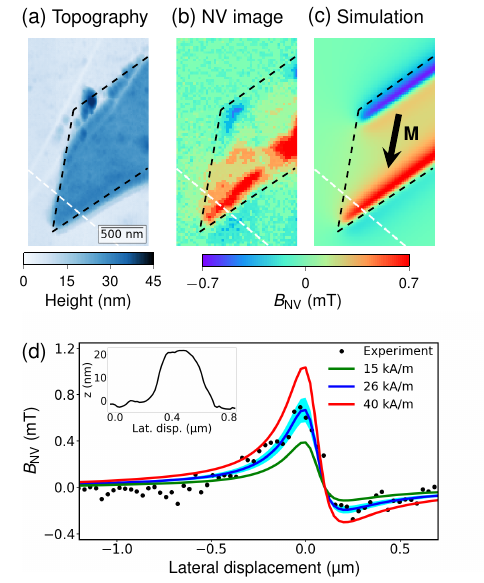}
		\caption{(a) AFM image of a 20-nm thick \CT flake. (b) Corresponding map of the magnetic field component $B_{\rm NV}$. (c) Simulated map of $B_{\rm NV}$ for a uniform in-plane magnetization $\mathbf{M}$ (black arrow) with an azimuthal angle $\phi_{M}=-100^{\circ}$ in the $(x,y)$ plane and a norm $M=26$~kA/m. The black dashed line indicate the shape of the flake used for the simulation. (d) Fitting of an experimental line profile (black markers) with the magnetic simulation (blue solid line). The linecuts are indicated by the white dashed lines in (b) and (c). A magnetization $M=26 \pm 4.0$~kA/m is obtained, the uncertainty being illustrated by the blue shaded area. The inset shows a linecut across the flake (white dashed lines in (a)).}
		\label{fig:simu_fullb}
	\end{figure}
	
The room-temperature magnetization measured in the exfoliated \CT flake is almost five times smaller than the one obtained in the bulk crystal. This observation could be explained by a degradation of the sample surface through oxidation processes, leading to a thinner effective magnetic thickness. However, recent experiments relying on X-ray magnetic circular dichroism coupled to photoemission electron microscopy (XMCD-PEEM) have shown that oxidation is limited to the first outer layers of the \CT flake~\cite{Coraux2020}. Considering a $35$-nm thick \CT flake, surface oxidation can thus be safely neglected, and cannot explain the observed reduction of the magnetization. This effect is rather attributed to a decrease of the Curie temperature in exfoliated flakes, a phenomenon that has been observed in other vdW magnets such as  Fe$_3$GeTe$_2$ below 5-10 nm thicknesses~\cite{FGT2018,Tan2018}, and in more traditional ferromagnetic thin films below few nanometers thicknesses~\cite{PhysRevB.49.3962,PhysRevLett.68.1208}. The thickness marking the crossover from a thin film (2D-like) to a bulk magnetism (3D) is expected to be strongly material-dependent and cannot be predicted a priori in the case of CrTe$_2$ \cite{Gibertini2019}. However, for a $35$-nm thick flake, bulk-like magnetism is a reasonable assumption and the magnetization's amplitude should not be altered by the film thickness. In this thickness regime, the main parameter altering the magnetization's amplitude is temperature, and how close or far it is from~$T_{\rm c}$. The estimation of the reduction in Curie temperature was tentatively inferred by translating the Curie law measured for a bulk \CT crystal [Fig.~1(a)] in order to reach $M\sim27$~kA/m at room temperature. This is achieved for a reduction of $T_{\rm c}$ by $\sim20$~K. This value is in line with a recent study, which estimates a Curie temperature around $300$~K for CrTe$_2$ flakes with a few tens of nanometers thickness~\cite{suncrte22019}.

To support this finding, a similar analysis was performed for a 20-nm thick flake [Fig.~\ref{fig:simu_fullb}(a)]. Here, a stray magnetic field is mainly detected along the bottom edge of the flake. On the opposite edge, the stray field distribution is quite inhomogeneous [Fig.~\ref{fig:simu_fullb}(b)]. This observation is attributed to damages of the flake, which can be observed in the AFM image. It is difficult to describe the corresponding complex thickness variations, and hence to take them into account in the simulations. These height variations seem to correspond to several CrTe$_2$ islands, whose very small sizes could make them more prone to oxidation than larger flakes, and which may exhibit random magnetization orientations. We hence disregarded the thickness variations in our structural model and the magnetic calculation was performed for a flake with uniform thickness, which is a good approximation for the bottom and left edges of the flake. A simulation of the stray field distribution for a magnetization with an azimuthal angle $\phi_{M}=-100\pm10^{\circ}$ reproduces fairly well the experimental results [Fig.~\ref{fig:simu_fullb}(c)] and a quantitative analysis of line profiles across the bottom edge of the flake leads to $M=26 \pm 4.0$~kA/m~[Fig.~\ref{fig:simu_fullb}(d)], a similar value to that obtained for the 35-nm thick flake. This observation indicates that the magnetization is not significantly modified for thicknesses lying in the few tens of nanometers range. A more in-depth analysis of the dependence of magnetization on thickness could not be carried out at this stage, given the very low probability to obtain thin \CT flakes of homogeneous thickness through mechanical exfoliation.
\section{Conclusion}

To conclude, we have used quantitative magnetic imaging with a scanning-NV magnetometer to demonstrate that exfoliated \CT flakes with thicknesses down to $20$~nm exhibit an in-plane ferromagnetic order at room temperature with a typical magnetization in the range of $M\sim 27$~kA/m. These results make \CT a unique system in the growing family of vdW ferromagnets, because it is the only material platform known to date which offers an intrinsic in-plane magnetization and a $T_{\rm c}$ above room temperature in thin flakes. These properties might offer several opportunities for studying magnetic phase transition in 2D-$XY$ systems~\cite{Parkin2020} and to design spintronic devices based on vdW magnets. The next challenge will be to assess if the ferromagnetic order is preserved at room temperature in the few layers limit.

\section*{Acknowledgements}
The authors warmly thank J.~Vogel and M.~Núñez-Regueiro for fruitful discussions. This research has received funding from the European Union H2020 Program under Grant Agreement No. 820394 (ASTERIQs), the DARPA TEE program, and the Flag-ERA JTC 2017 project MORE-MXenes. A.F. acknowledges financial support from the EU Horizon 2020 Research and Innovation program under the Marie Sklodowska-Curie Grant Agreement No. 846597 (DIMAF).

\section*{Appendix A: Magnetic field simulation}

\begin{figure*}[t]
		\centering
		\includegraphics[width = 13cm]{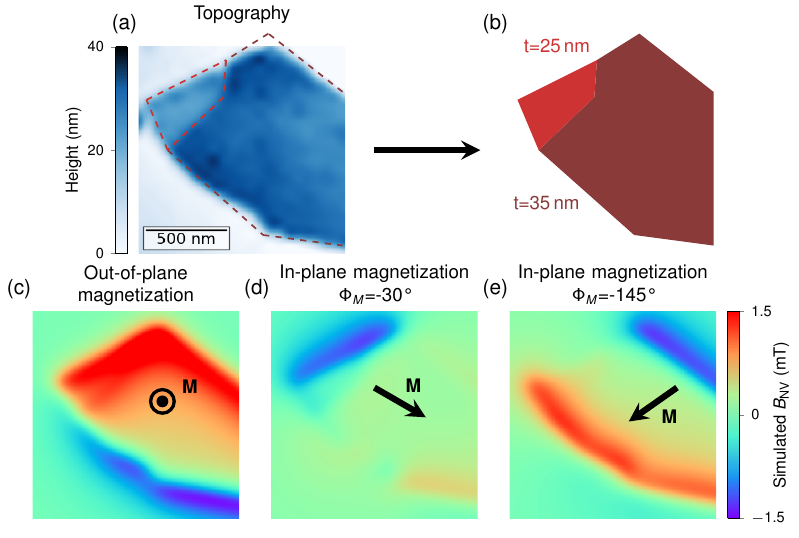}
		\caption{(a) AFM image of \CT flake shown in Fig. 3. (b) Geometry of the flake used for the magnetic calculation, which includes the thickness step observed in the AFM imafge. (c-e) Calculated maps of $B_{\rm NV}$ for a uniform magnetization $\mathbf{M}$ pointing out-of plane (c), and in-plane with an azimuthal angle $\phi_{M}=-30^{\circ}$ (d) or $\phi_{M}=-145^{\circ}$ (e). The norm of the magnetization is fixed to $M=27$~kA/m.}
		\label{fig:supSimu}
	\end{figure*}
	
As indicated in the main text, thickness variations within the flake can result in a magnetization pattern with a non-zero divergence which produces a stray magnetic field. When possible, these variations were taken into account in the magnetic calculation. This is illustrated in Fig.~\ref{fig:supSimu}(a,b), where the geometry of the flake used for the calculation includes a thickness step observed in the topography image. In Fig.~\ref{fig:supSimu}(c-e), we show the magnetic field distributions produced at a distance $d_{\rm NV}=80$~nm for three different magnetization orientations of the flake. First considering an out-of-plane magnetization [Fig.~\ref{fig:supSimu}(c)], the simulated magnetic field distribution does not agree with the experimental data shown in Fig. 3. This confirms that the magnetization in lying in the plane of the \CT flake, {\it i.e.} perpendicular to the $c$ axis as obtained in the bulk crystal. Simulations performed for a uniform in-plane magnetization for two different values of the azimuthal angle $\phi_{M}$ in the $(x,y)$ plane are shown in Fig.~\ref{fig:supSimu}(d,e). A comparison with experimental data allows the identification of the magnetization orientation. 
	
\section*{Appendix B: Analysis of uncertainties}

The norm of the magnetization $M$ is obtained by fitting line profiles of the measured stray field distribution with the result of the magnetic calculation [see~Fig.~3(d)]. In this section, we analyze the uncertainty of this measurement by using the methodology described in Ref.~\cite{gross_real-space_2017}. The uncertainties result (i) from the fitting procedure itself and (ii) from those on the parameters $p_i=\{d_{\rm NV},\theta_{\rm NV},\phi_{\rm NV},t,\phi_{M}\}$ which are all involved in the magnetic calculation. In the following, the parameters $p_i$ are expressed as $p_i=\bar{p_i}+\sigma_{p_i}$, where $\bar{p_i}$ denotes the nominal value of parameter $p_i$ and $\sigma_{p_i}$ its standard error. These parameters are independently evaluated as follows:
\begin{itemize}
\item[\hl{$\bullet$}] \ The probe-to-sample distance $d_{\rm NV}$ is inferred through a calibration measurement, following the procedure described in Ref.~\cite{Hingant2015}, leading to $d_{\rm NV}=80\pm 10$~nm.
\item[$\bullet$] \ The NV defect quantization axis is measured by recording the ESR frequency as a function of the amplitude and orientation of a calibrated magnetic field, leading to spherical angles $(\theta_{\rm NV}=58\pm2^{\circ},\phi_{\rm NV}=103\pm2^{\circ})$ in the laboratory frame of reference $(x,y,z)$ [see~Fig.~1(b)]. 
 \item[$\bullet$] \ The thickness $t$ of the \CT flake is extracted from line profiles of the AFM image with an uncertainty of $\pm 2$~nm. 
  \item[$\bullet$] \ The azimuthal angle of the in-plane magnetization $\phi_{M}$ is obtained through the comparison between experimental data and simulated magnetic field maps.
 \end{itemize}
 
We first evaluate the uncertainty of the fitting procedure. To this end, the line profile is fitted with the result of the magnetic calculation while fixing all the parameters $p_i$ to their nominal values $\bar{p_{i}}$, leading to $M=26.9\pm0.7$~kA/m. The relative uncertainty linked to the fitting procedure is therefore given by $\epsilon_{\rm fit}=3 \%$. We note that the intrinsic accuracy of the magnetic field measurement is in the range of $\delta B_{\rm NV}\sim$~$5 \ \mu$T. The resulting uncertainty can be safely neglected.

In order to estimate the relative uncertainty $\epsilon_{p_{i}}$ introduced by each parameter $p_i$, the fit was performed with one parameter $p_i$ fixed at $p_i=\bar{p_{i}} \pm \sigma_{p_{i}}$, all the other parameters remaining fixed at their nominal values. The corresponding fit outcomes are denoted $M(\bar{p_{i}} + \sigma_{p_{i}})$ and $M(\bar{p_{i}} - \sigma_{p_{i}})$. The relative uncertainty $\epsilon_{p_{i}}$ introduced by the errors on parameter $p_i$ is then finally defined as
\begin{equation} \label{partial_uncert}
\epsilon_{p_{i}}=\frac{M(\bar{p_{i}} + \sigma_{p_{i}})-M(\bar{p_{i}} - \sigma_{p_{i}})}{2M(\bar{p_{i}} )} \ .
\end{equation}
This analysis was performed for each parameter $p_i$. The cumulative uncertainty $\epsilon$ is finally given by 
\begin{equation}
\epsilon=\sqrt{\epsilon^{2}_{\rm fit}+\sum_i \epsilon_{p_i}^2} \ ,
\label{EqUncert}
\end{equation}
where all errors are assumed to be independent.

A summary of the uncertainties is given in Table I. We obtain $M=27\pm 4$~kA/m for the flake shown in Fig. 3 and $M=26\pm 4$~kA/m for the flake shown in Fig.~4.

\begin{table}[h!]
\label{tabUncert}
\caption{Analysis of uncertainties in the measurement of the magnetization}
\begin{center}
{(a) \CT flake shown in Figure 3}\\
\vspace{0.5cm}
\begin{tabular}{|c|c|c|c|}
\hline
parameter $p_i$ &  \ nominal value $\bar{p_i}$ \ & \ uncertainty $\sigma_{p_i}$ \ & $ \ \epsilon_{p_i}(\%) \ $   \\
\hline
 $d_{\rm NV}$ & 80 nm & 10 nm & 8 \\
 $\theta_{\rm NV}$ & $58^\circ$ & $2^\circ$ & 7  \\
 $\phi_{\rm NV}$ & $103^\circ$ & $2^\circ$ & 2 \\
 $t$ & $35$~nm & $2$~nm & 6  \\
 $\phi_M$ & $-145^\circ$ & $5^\circ$ & 5  \\
 \hline
  \multicolumn{3}{|c|}{} & \\
 \multicolumn{3}{|c|}{$\epsilon=\sqrt{\epsilon^{2}_{\rm fit}+\sum_i \epsilon_{p_i}^2}$} & 14 \\
  \multicolumn{3}{|c|}{} & \\
   \hline
\end{tabular}
\end{center}

\begin{center}
{(b) \CT flake shown in Figure 4}\\
\vspace{0.5cm}
\begin{tabular}{|c|c|c|c|}
\hline
parameter $p_i$ &  \ nominal value $\bar{p_i}$ \ & \ uncertainty $\sigma_{p_i}$ \ & $ \ \epsilon_{p_i}(\%) \ $   \\
\hline
 $d_{\rm NV}$ & 80 nm & 10 nm & 8 \\
 $\theta_{\rm NV}$ & $58^\circ$ & $2^\circ$ & 7  \\
 $\phi_{\rm NV}$ & $103^\circ$ & $2^\circ$ & 2 \\
 $t$ & $24$~nm & $2$~nm & 8  \\
 $\phi_M$ & $-100^\circ$ & $10^\circ$ & 12  \\
 \hline
  \multicolumn{3}{|c|}{} & \\
 \multicolumn{3}{|c|}{$\epsilon=\sqrt{\epsilon^{2}_{\rm fit}+\sum_i \epsilon_{p_i}^2}$} & 17 \\
  \multicolumn{3}{|c|}{} & \\
   \hline
\end{tabular}

\label{tab0}
\end{center}
\end{table}

	\bibliography{CrTe2_biblio}
	\bibliographystyle{apsrev4-1}
	
\end{document}